# FERROELECTRIC AND INCIPIENT FERROELECTRIC PROPERTIES OF A NOVEL $Sr_{9-x}Pb_xCe_2Ti_2O_{36}$ (x=0-9) CERAMIC SYSTEM

Stanislav Kamba[1]*, Maxim Savinov[1], František Laufek[2], Ondřej Tkáč[1], Christelle Kadlec[1], Sergiy Veljko[1], Elizabeth J. John[1], Ganesanpotti Subodh[3], Mailadil T. Sebastian[3], Mariana Klementová[4], Viktor Bovtun[1], Jan Pokorný[1], Veronica Goian[1], Jan Petzelt[1]

[1]Institute of Physics ASCR, Na Slovance 2, 18221 Prague 8, Czech Republic

[2]Czech Geological Survey, Geologická 6, 152 00 Prague 5, Czech Republic

[3]National Institute for Interdisciplinary Science and Technology, Trivandrum 695019, India

[4]Institute of Inorganic Chemistry, ASCR, v.v.i., 250 68 Husinec-Řež 1001, Czech Republic

*Corresponding author:
Stanislav Kamba,
Phone: +420 266052957
Fax: +420286890527
E-mail: kamba@fzu.cz

Received date:

**Title running head:** Novel ferroelectric system $Sr_{9-x}Pb_xCe_2Ti_2O_{36}$ (x=0-9).

**Brief summary:**

Novel ferroelectric $Sr_{9-x}Pb_xCe_2Ti_{12}O_{36}$ (x=0-9) ceramics were prepared. Structural phase transitions from the paraelectric trigonal $R\bar{3}c$ phase to the ferroelectric monoclinic $Cc$ phase were discovered in ceramics for x≥3. Ferroelectric soft mode was observed in THz and infrared spectra, so the structural phase transitions are of the displacive type.

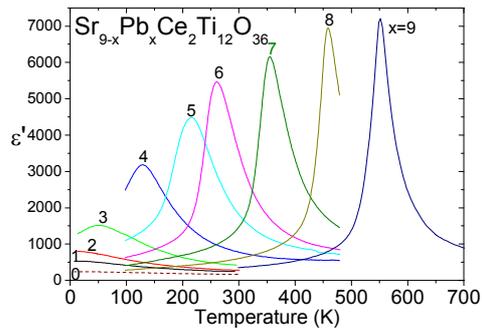

**Abstract**

$Sr_{9-x}Pb_xCe_2Ti_{12}O_{36}$ system is derived from the perovskite $SrTiO_3$ and its chemical formula can be written as $(Sr_{1-y}Pb_y)_{0.75}Ce_{0.167}TiO_3$. We investigated dielectric response of $Sr_{9-x}Pb_xCe_2Ti_{12}O_{36}$ ceramics (x = 0-9) between 100 Hz and 100 THz at temperatures from 10 to 700 K using low- and high-frequency dielectric, microwave (MW), THz and infrared spectroscopy. We revealed that $Sr_9Ce_2Ti_{12}O_{36}$ is an incipient ferroelectric with the $R\bar{3}c$ trigonal structure whose relative permittivity ε' increases from 167 at 300 K and saturates near 240 below 30 K. The subsequent substitution of Sr by Pb enhances ε' to several thousands and induces a ferroelectric phase transition to monoclinic $Cc$ phase for x≥3. Its critical temperature $T_C$ linearly depends on the Pb concentration and reaches 550 K for x=9. The phase transition is of displacive type. The soft mode frequency follows the Barrett formula in samples with x≤2 (typical for quantum paraelectrics) and the Cochran formula for x≥3. The MW dispersion is lacking and quality factor Q is high in samples with low Pb concentration, although the permittivity is very high in some cases. However, due to the lattice softening, the temperature coefficient of the permittivity is rather high. The best MW quality factor was observed for x=1: Q*f=5800 GHz and ε'=250. Concluding, the dielectric properties of $Sr_{9-x}Pb_xCe_2Ti_{12}O_{36}$ are similar to those of $Ba_{1-x}Sr_xTiO_3$ so that this system can be presumably used as an alternative for MW devices or capacitors.

## 1. Introduction

Nowadays, tunable dielectric materials are widely used in microwave (MW) devices such as filters, phase shifters and antennas.[1,2,3] Ferroelectric (FE) materials are used for this application as they have high dielectric permittivity ε' close to the FE phase transition temperature $T_C$ which can be tuned by an electric field[1-3]. However, in general, materials in the FE phase exhibit higher dielectric losses due to the absorption of FE domain walls.[1] Therefore it is more appropriate to use materials in the paraelectric phase close but above $T_C$. Moreover, it is necessary to use materials with a displacive FE phase transition driven by an unstable lattice vibration (called FE soft optic phonon or shortly soft mode - SM) lying in the THz range. In the paraelectric phase, the permittivity of such systems is completely described by the contribution of polar optic phonons. Therefore there is no dielectric relaxation below the phonon frequencies and the dielectric losses in the MW range are low being caused mainly by intrinsic multiphonon absorption and possibly extrinsic contributions stemming from the defects.[4,5] However, most of the ferroelectrics undergo an order-disorder type of the phase transition. In this case the soft excitation is not an optic phonon, but a soft dielectric relaxation due to hopping of some disordered sublattice among several equivalent positions.[6] Such relaxations lie typically in the MW region, therefore the MW dielectric losses are high. Such materials are not suitable for MW applications.

SrTiO$_3$, CaTiO$_3$ and KTaO$_3$ exhibit high permittivity and no MW dielectric dispersion (i.e. low MW losses). These materials belong to the so called incipient ferroelectrics or quantum paraelectrics,[7] which are characterized by increasing permittivity on cooling due to the softening of the lowest frequency polar optical phonon. The SM frequency $\omega_{SM}$ at high temperatures follows the Cochran law $\omega_{SM} = \sqrt{A(T - T_{cr})}$ like in usual ferroelectrics,[6] but the FE phase is never reached, because the hypothetical critical temperature $T_{cr}$ lies below 0 K (in CaTiO$_3$ and KTaO$_3$). Sometimes $T_{cr} > 0$ K ($T_{cr} \sim +37$ K in SrTiO$_3$) but quantum fluctuations prevent the appearance of the FE order and permittivity saturates at low temperatures. Such materials are called quantum paraelectrics being a subset of incipient ferroelectrics. The quantum fluctuations prevent excessive phonon softening, and therefore the soft mode frequency at low temperature does not follow the Cochran law and saturates at a final low value according to the modified Barrett formula[8,9]

$$\omega_{SM} = \sqrt{A\left[\left(\frac{T_1}{2}\right)\coth\left(\frac{T_1}{2T}\right) - T_{cr}\right]}. \qquad (1)$$

$T_1$ marks the temperature below which the quantum fluctuations start to play a role while above this temperature the SM follows the Cochran law. $T_{cr}$ corresponds to the extrapolated (hypothetical) critical temperature from the Cochran law.

All the above mentioned incipient ferroelectrics exhibit very high permittivities ($\varepsilon'$ = $10^3$ - $10^4$) at low temperatures, but their room temperature values are only 150-300.[10] Their dielectric losses are rather low, but the temperature coefficient of the resonance frequency $\tau_f$ (which characterizes the temperature dependence of the permittivity: $\tau_f \cong -\frac{d\varepsilon'}{2\varepsilon' dT}$) is very high. There were many attempts to dope or process solid solutions of incipient ferroelectrics with other dielectrics. It was possible to reduce $\tau_f$, but $\varepsilon'$ was simultaneously drastically reduced. Only in the case of solid solutions with ferroelectrics (like $BaTiO_3$ or $PbTiO_3$)[11,12] it was possible to enhance permittivity at room temperature, but simultaneously $\tau_f$ and the dielectric loss $\varepsilon''$ dramatically rose up. In spite of this, $Ba_{1-x}Sr_xTiO_3$ is nowadays the most used material in MW tunable devices thanks to its high and tunable $\varepsilon'$ and its moderately low $\varepsilon''$. Nevertheless, a search for new MW materials with low $\varepsilon''$ and $\tau_f$ but high $\varepsilon'$ tunable by external electric field, still continues.

Very recently Subodh et al.[13] discovered a new $Sr_{2+n}Ce_2Ti_{5+n}O_{15+3n}$ (n≤10) homologous series, alternatively written as $Sr_{1-3x/2}Ce_xTiO_3$ (x≤0.40). This ceramics system shows $\varepsilon'$ ranging from 113 to 185 and a high value of Q*f between 6000 and 11000 GHz (Q*f parameter, where f marks the measuring frequency and the microwave quality factor is defined as Q=ε'/ε'', is usually used for comparison of the MW losses of different materials). Moreover, $\tau_f$ values are five times smaller than those commonly observed in SrTiO$_3$. Moreira et al.[14] investigated optical phonon spectra of Sr$_{1-3x/2}$Ce$_x$TiO$_3$ (x≤0.40) by means of infrared (IR) and Raman spectroscopy and have shown that the MW ε' values are completely determined by contributions of polar optical phonons, i.e. no dielectric relaxation exists below the phonon frequencies. It is compatible with the high Qxf values observed in this dielectric system. Subodh et al.[13] and Moreira et al.[14] observed a cubic symmetry of Sr$_{2+n}$Ce$_2$Ti$_{5+n}$O$_{15+3n}$ (n≤10) derived from SrTiO$_3$, but very recent and detailed structural studies of Ubic et al.[15] revealed the trigonal $R\bar{3}c$ space group at room temperature (RT).

In the present article we report the structural phase diagram and MW dielectric properties of Sr$_{9-x}$Pb$_x$Ce$_2$Ti$_{12}$O$_{36}$ ceramics with x=0-9. Sr$_9$Ce$_2$Ti$_{12}$O$_{36}$ was selected for Pb substitution as it has relatively high ε' and low ε'' among other Sr$_{2+n}$Ce$_2$Ti$_{5+n}$O$_{15+3n}$ ceramics and moreover four times lower $\tau_f$ than in SrTiO$_3$. We will show that Pb substitution dramatically enhances ε' and induces FE phase transition. The nature of the phase transition will be investigated by means of THz and IR spectroscopy and it will be shown that the SM frequency strongly depends on Pb concentration and temperature. THz spectra are sensitive mainly to the intrinsic dielectric loss contributions[4] and since such losses can be linearly extrapolated into the MW frequency range[1,4], we will estimate also the contributions of intrinsic losses to the measured MW loss. The phase diagram will be determined based on X-ray diffraction (XRD) and selected area electron diffraction (SAED) data combined with IR and Raman scattering spectra.

## 2. Experimental

$Sr_{9-x}Pb_xCe_2Ti_{12}O_{36}$ (x=0 to 9) ceramics were prepared by the solid-state ceramic route. High purity $SrCO_3$, PbO and $TiO_2$ (99.9+%, Aldrich Chemical Company, Inc, Milwaukee, WI, USA) and $CeO_2$ (99.99%, Indian Rare Earth Ltd, Udyogamandal, India) were used as the starting materials. Stoichiometric amounts of powder mixtures were ball-milled in distilled water medium using yttria-stabilized zirconia balls in a plastic container for 24h. The slurry was dried, ground and the materials were calcined in Pt crucibles, depending on the Pb content at temperature ranging from 950 to 1100 $^o$C for 4h. The calcined material was ground into a fine powder and divided into different batches for optimizing the sintering temperature. Around 4-wt% of polyvinyl alcohol (PVA) (molecular weight 22000, BDH Lab Suppliers, England) was added to the dried powders and again ground into fine powder. Cylindrical pucks of about 4-5 mm height and 11 mm diameter were made by applying a pressure of 100 MPa. The sintering temperatures were optimized for the best density and depending on the composition of Pb the samples were sintered at temperatures 1100 to 1350 $^o$C for 2h. The pellets were muffled in the same powder in order to prevent escape of the volatile Pb.

The XRD patterns of the samples were taken in Bragg-Brentano geometry by the X´Pert Pro PANalytical diffractometer equipped with X´Celerator detector using Cu-Kα radiation. Unit cell parameters, atomic positions, and isotropic displacement parameters were refined by means of Rietveld method using the FullProf program.[16] Transmission electron microscopy of $Sr_9Ce_2Ti_{12}O_{36}$ and $Pb_9Ce_2Ti_{12}O_{36}$ ceramics was carried out on a JEOL JEM 3010 microscope operated at 300 kV (LaB6 cathode, point resolution 1.7Å). Powder samples were dispersed in ethanol and the suspension was treated in ultrasound for 10 minutes. A drop of very dilute suspension was placed on a holey-carbon-coated copper grid and allowed to dry by evaporation at RT.

The low-frequency dielectric response was investigated between 100 Hz and 1 MHz from 10 to 700 K using an impedance analyzer HP 4192A. A high-frequency response (1 MHz – 1.8 GHz) was obtained by means of the impedance analyzer Agilent 4291B with Novocontrol BDS 2100 coaxial sample cell and a Sigma System M18 temperature chamber (operating range 100 – 570 K). At MW, cylindrical samples were measured as dielectric resonators in the shielding cavity between 100 and 370 K using the network analyzer Agilent E8364B. Their resonance frequencies (0.9-3.2 GHz) depend on the values of $\varepsilon'$ and the size of the samples (diameter 9 mm, height 4.0-4.7 mm). Measurements at THz frequencies from 0.2 to 2.5 THz (7 – 80 cm$^{-1}$) were performed at RT using a time-domain THz spectrometer. THz spectra of $Pb_9Ce_2Ti_{12}O_{36}$ ceramics were also investigated at elevated temperatures up to 900 K, $Sr_9Ce_2Ti_{12}O_{36}$ ceramics were measured down to 10 K. IR reflectivity spectra were obtained using a Fourier transform IR spectrometer Bruker IFS 113v in the frequency range 20 - 3000 cm$^{-1}$ (0.6 – 90 THz). $Sr_{9-x}Pb_xCe_2Ti_{12}O_{36}$ samples with x = 0, 3, 6 and 9 were investigated between 10 and 900 K. For details about our THz and IR technique see Ref. 9. RT Raman spectra were measured in backscattering geometry using a Raman microscope (Renishaw RM 1000) equipped with a grating filter (NExT) to enable Raman shift measurements down to 20 cm$^{-1}$. An Ar-laser line (514 nm) was used for the excitation.

### 3. Results and Discussion

#### (a) Phase diagram of $Sr_{9-x}Pb_xCe_2Ti_{12}O_{36}$ (x=0-9) ceramics.

Fig. 1 shows the RT X-ray powder diffraction patterns of all $Sr_{9-x}Pb_xCe_2Ti_{12}O_{36}$ (x=0-9) ceramics. One can see that for x > 7, some of the diffraction lines split due to the change of crystal symmetry. Latter we will demonstrate, based on the dielectric data, that the new phase is FE. We found that all the patterns from x = 0 to 6 resemble those of a typical cubic perovskite structure ($Pm\bar{3}m$) as reported in Refs. 13,14 for the parent compound $Sr_9Ce_2Ti_{12}O_{36}$. However, Ubic et al.[15] recently reported the electron and neutron diffraction of $Sr_{1-3x/2}Ce_xTiO_3$ (x≤0.40) ceramics, which revealed superlattice reflections (not seen in XRD) and therefore they supposed that $Sr_9Ce_2Ti_{12}O_{36}$ has non-cubic supercell caused by antiphase tilting of oxygen octahedra. Its space group should be $R\bar{3}c$, corresponding to the a⁻a⁻a⁻ tilt system.[15]

Refinement of our XRD patterns of samples with x ≤ 6 was successful in all cubic, tetragonal and trigonal structures. For determining the structure we also performed SAED measurements of $Sr_9Ce_2Ti_{12}O_{36}$ and observed the same satellites as reported by Ubic et al.[15] (see Fig. 2a), supporting only the trigonal $R\bar{3}c$ structure. The lattice parameters $a$ = 5.5092(1) Å and $c$ = 13.4989(4) Å obtained from XRD correspond very well with those reported by Ubic et al.[15]

The question arises what is the crystal symmetry of the FE phase in $Sr_{9-x}Pb_xCe_2Ti_{12}O_{36}$? It is natural to assume that it should be a subgroup of $R\bar{3}c$. Rietveld refinement of $Pb_9Ce_2Ti_{12}O_{36}$ phase in trigonal $R3c$ structure was not successful (see inset of Fig. 3), however the monoclinic $Cc$ ($C_S^4$) structure with lattice parameters a = 9.7342(5) Å, b = 5.5330(6) Å, c = 5.5343(6) Å and β = 124.78(3)° corresponds well to the powder diffraction (Fig. 3) and SAED data (Fig. 2). This monoclinic structure model was derived from the $Sr_9Ce_2Ti_{12}O_{36}$ structure by means of appropriate crystallographic transformation (i.e. Wyckoff splitting) from $R\bar{3}c$ space group to its subgroup $Cc$ using PowderCell program.[17] The final cycles of Rietveld refinement of $Pb_9Ce_2Ti_{12}O_{36}$ converged to residual factors $R_p$ = 6.54, $R_{wp}$ = 8.76 and $R_{Bragg}$ = 5.61 %; crystallographic data of $Pb_9Ce_2Ti_{12}O_{36}$ are summarized in Table 1. It should be noted, that the positions of oxygen atoms in this structure cannot be accurately refined from the XRD data, because of high scattering contrast among Pb and O atoms. Nonetheless, the determined structure is quite informative, especially with respect to lowering symmetry in the whole system. The slight increase of the lattice parameter with Pb content was observed due to a larger ionic radius of Pb (1.49 Å) compared to 1.44 Å of Sr.[18] It is worth noting that the elementary perovskite subcell parameters $a_p = c_m/\sqrt{2} = 3.913$ Å, $b_p = b_m/\sqrt{2} = 3.912$ Å, $c_p = 0.5[a_m^2 + c_m^2 - 2 a_m c_m \cos(\pi-\beta)]^{1/2} = 3.998$ Å of $Pb_9Ce_2Ti_{12}O_{36}$ phase indicate pseudotetragonal relationship ($a_p \approx b_p, \neq c_p$), which can be also observed from SAED patterns (Fig. 2).

Below we will demonstrate how the IR and Raman spectra correspond to the crystal structures in the paraelectric and FE phase. However, let us first discuss the dielectric properties of the investigated $Sr_{9-x}Pb_xCe_2Ti_{12}O_{36}$ ceramics as a function of temperature, which enable us to determine the temperature and type of the FE transitions in relation to the Pb concentration. We measured the temperature dependence of the dielectric response between 100 Hz and 3.2 GHz. Representative curves for all ten samples (x = 0 - 9) are shown in Fig. 4. No dielectric dispersion, i.e. no change of permittivity with frequency, was observed. Therefore we plot the curves for each sample only at a single frequency. Samples with x ≤ 3 and x = 9 have anomalies out of the temperature region of our high-frequency setup (100 – 500 K), therefore curves for such compositions are plotted at 100 kHz. Low-frequency data below 1 MHz show small (less than 3%) dielectric dispersion, best seen near temperatures of permittivity maxima (not shown here). The peak of ε'(T) is not shifting with frequency and no dispersion was observed in the high-frequency range above 1 MHz so that the samples show no relaxor properties. The low-frequency dispersion originates probably from the point defects (vacancies etc.) or grain boundaries. Maxima of ε'(T) correspond to the temperatures $T_C$ of the FE transitions. The phase diagram seen in Fig. 5 clearly shows that $T_C$ linearly increases with Pb concentration. The FE hysteresis loop taken at RT in $Pb_9Ce_2Ti_{12}O_{36}$ shows a spontaneous polarization of 2.5 μC/cm² (see inset in Fig. 5). We also note that piezoelectricity with $d_{33}$ = 65 pC.N$^{-1}$ was recently reported in this material.[19]

RT MW dielectric properties are summarized in Fig. 6. MW dielectric resonances were obtained in all the $Sr_{9-x}Pb_xCe_2Ti_{12}O_{36}$ ceramics except for the x = 6 sample with maximum ε' whose value in Fig. 6 is from the coaxial high-frequency experiment since it exhibits high losses preventing the MW resonance. One can see that the best Q*f was obtained for the x=1 sample, where Q*f = 5800 GHz and ε' = 250. However, its $\tau_f$ = 920 ppm/K is too high for MW applications, although it is almost twice smaller than in $SrTiO_3$.

**(b) Room-temperature IR and Raman spectra and the crystal structure**

IR and Raman spectra can help in refinement of the crystal structure and, moreover, in the case of displacive phase transition a SM should be seen. Moreover, dielectric loss calculated from the fits of IR and THz spectra can be extrapolated to the MW range and one can estimate the contribution of intrinsic and extrinsic loss to the experimentally observed MW loss.[4,5] We measured RT IR reflectivity, THz transmission and Raman scattering spectra of all the samples. IR spectra of selected compounds are shown in Fig. 7, Raman spectra of all samples are presented in Fig. 8.

In order to obtain polar phonon parameters, IR and THz spectra were fitted simultaneously using the generalized-oscillator model with the factorized form of the complex permittivity[6]

$$\varepsilon^*(\omega) = \varepsilon_\infty \prod_j \frac{\omega_{LOj}^2 - \omega^2 + i\omega\gamma_{LOj}}{\omega_{TOj}^2 - \omega^2 + i\omega\gamma_{TOj}} \qquad (2)$$

where $\omega_{TOj}$ and $\omega_{LOj}$ denote the transverse and longitudinal frequency of the j-th polar phonon, respectively, and $\gamma_{TOj}$ and $\gamma_{LOj}$ their damping constants. The high-frequency permittivity $\varepsilon_\infty$ resulting from electronic absorption processes was obtained from the frequency-independent reflectivity above the phonon frequencies. $\varepsilon^*(\omega)$ is related to the IR reflectivity R(ω) by

$$R(\omega) = \left| \frac{\sqrt{\varepsilon^*(\omega)} - 1}{\sqrt{\varepsilon^*(\omega)} + 1} \right|^2. \tag{3}$$

Real and imaginary parts of the complex dielectric spectra $\varepsilon^*(\omega)=\varepsilon'(\omega)-i\varepsilon''(\omega)$ obtained from the fits of the RT IR and THz spectra together with the dotted THz data are shown in Fig. 9. Polar phonon frequencies are plotted versus the Pb concentration in Fig. 10. One can see that due to the change of crystal symmetry in samples for $x \geq 7$ many new modes are activated in both IR and Raman spectra since they are in the FE phase at RT, while the $x \leq 6$ samples are in the paraelectric phase. Inset in Fig. 10 clearly shows a lowest-frequency mode softening, whose frequency remarkably depends on the Pb concentration exhibiting a minimum for x=6 (corresponding $T_C$ = 260 K is close to the RT). Softening of the mode is manifested in Fig. 9 by a shift of the peak in $\varepsilon''(\omega)$ to lower frequencies as well as by increase of the static permittivity with the maximum value for x=6. The oscillator strength $f_{SM}=\Delta\varepsilon_{SM}\cdot\omega_{TO}^2$ of the SM is the strongest out of all phonons in all the samples, therefore this mode could be assigned to the vibrations of the Ti cation against the oxygen octahedron, so called Slater mode.[20] This is not surprising in Sr-rich samples, because Slater vibration is the SM also in the related $SrTiO_3$. However, it was not expected in Pb rich samples, because the eigenvector of the SM in the related $PbTiO_3$ describes predominantly the Pb vibrations against the oxygen octahedron (Last mode[21]). The statement that SM is the Slater mode in all x=1-9 samples is also supported by our XRD data of $Pb_9Ce_2Ti_{12}O_{36}$, which show that the Pb cations remain unshifted in the FE phase and only oxygen and predominantly Ti cations shift from their paraelectric positions. For more details how to distinguish Last and Slater mode from the IR spectra see the review by Hlinka et al.[22]

Let us compare the number of observed phonons with the prediction of the factor group analysis (using the tables by Rousseau et al.[23]) for different suggested crystal symmetries for the $Sr_{9-x}Pb_xCe_2Ti_{12}O_{36}$ ceramics. In Table 2 we listed the symmetries of phonons and their IR and Raman (R) activities in acceptable crystal phases. Let us discuss first the spectra in the paraelectric phase. Their typical representative is the lead free $Sr_9Ce_2Ti_{12}O_{36}$, whose IR and Raman spectra were recently investigated by Moreira et al.[14] Our spectra correspond very well with the published ones. They exhibit three distinct reflection bands, which could correspond to previously suggested cubic $Pm\bar{3}m$ space group[13,14] with 3 triply degenerate polar phonons of $F_{1u}$ symmetry.[24] However, our detailed fits of the IR spectra revealed 6 modes in all the paraelectric samples with x ≤ 6 (see Fig. 10). Moreira et al.[14] suggested that the low-temperature structure of $Sr_9Ce_2Ti_{12}O_{36}$ could be tetragonal, in analogy with $SrTiO_3$, but they did not specify the space group. The tetragonal *P4/mmm* structure does not correspond to the satellites observed in SAED (Fig. 2) and the *I4/mcm* structure was excluded by Ubic et al., because some observed superlattice reflections are forbidden even in double diffraction for the *I4/mcm* group.[15]

According to Table 2, 8 and 12 IR active modes are expected in the trigonal $R\bar{3}c$ and tetragonal *I4/mcm* structure, respectively. Our 6 observed polar modes correspond better to the trigonal structure (the two missing modes can be overlapped by other stronger modes). Trigonal structure for the paralectric phase is also supported by Raman spectra. In Fig. 8 we see only a few very weak Raman bands in $Sr_9Ce_2Ti_{12}O_{36}$ and from Table 2, 5 Raman active phonons should be expected in the trigonal $R\bar{3}c$ crystal structure, while only 2 modes are expected in the *I4/mcm* structure. So, our IR and Raman spectra are in agreement with the result of XRD and SAED measurements suggesting the trigonal $R\bar{3}c$ structure for the paraelectric $Sr_9Ce_2Ti_{12}O_{36}$ ceramics. The $Sr_{9-x}Pb_xCe_2Ti_{12}O_{36}$ ceramics with x ≤ 6 show the same number of active phonons suggesting that their structure is the same. (We note that sharp Raman peak seen at 465 cm$^{-1}$ in the x ≥ 4 samples originates from the $CeO_2$ second phase[25] detected also by our XRD analysis).

The x ≥ 3 samples undergo a FE phase transition (see Figs. 4 and 5). According to our structural data, the FE phase is monoclinic with the *Cc* space group. Let us compare the factor group analysis performed for the *Cc* phase with our IR and Raman spectra of $Pb_9Ce_2Ti_{12}O_{36}$ with the highest $T_C$ where the modes activated in the FE phase are expected to be the strongest in the RT spectra. (Actually, from Fig. 10 one can see that $Pb_9Ce_2Ti_{12}O_{36}$ exhibits the highest number of polar modes. Similar situation holds in the Raman spectra – see Fig. 8.) We observed 14 polar phonons in the IR reflectivity spectra of $Pb_9Ce_2Ti_{12}O_{36}$ out of which the one near 380 cm$^{-1}$ probably corresponds to a polar phonon of $TiO_2$[26] identified as the second phase in our XRD analysis. So we can affirm 13 IR and 9 Raman active (see Fig. 8) modes, whereas the factor group analysis in Table 2 predicts 27 modes both IR and Raman active. The disagreement is rather large, but it can be understood by low strengths and overlapping of many newly activated modes which might not be resolved in the spectra.

We conclude that the IR and Raman spectra of $Sr_{9-x}Pb_xCe_2Ti_{12}O_{36}$ ceramics helped us in refinement of the crystal structures of this complex ceramic system. Although XRD and SAED analysis gave no unambiguous results for the paraelectric phase and both tetragonal *I4/mcm* and trigonal $R\bar{3}c$ symmetry were possible, our IR and Raman spectra supported the trigonal $R\bar{3}c$ phase. In the same way we supported the monoclinic FE phase with the space group *Cc*.

**(b) Dynamics of phase transitions**

How are the polar phonon frequencies dependent on temperature in the $Sr_{9-x}Pb_xCe_2Ti_{12}O_{36}$ ceramics? Can they explain the temperature dependence of MW and low-frequency permittivity? Are the phase transitions of order-disorder or displacive type? This will be discussed in the following subsection.

Fig. 11 shows temperature dependence of the IR reflectivity of $Sr_9Ce_2Ti_{12}O_{36}$ ceramics. One can see a small increase in the reflectivity and appearance of a new mode near 430 cm$^{-1}$ on cooling. Both effects are mostly due to a reduced phonon damping at low temperatures, but the rising of the low-frequency reflectivity below ~100 cm$^{-1}$ is caused by a small phonon softening, which fully accounts for the increase in the low-frequency ε' on cooling (see the inset in Fig. 11) and its saturation below ~30 K. The $Sr_{9-x}Pb_xCe_2Ti_{12}O_{36}$ samples with x≤2 exhibit qualitatively the same temperature dependence of reflectivity, since they do not undergo any phase transition at low temperatures. The value of the low-frequency reflectivity is slightly more increasing with rising x due to a lower SM frequency and higher value of ε' (Fig. 4).

Temperature dependence of ε' in the samples of 0≤x≤2 shown in Fig. 4 was successfully fitted with the Barrett formula[8]

$$\varepsilon' = \frac{M}{\frac{T_1}{2}\coth\left(\frac{T_1}{2T}\right) - T_{cr}} + \varepsilon_{B\infty}, \qquad (4)$$

where $T_{cr}$ and $T_1$ have the same meaning as in (Eq. 1), M is the Barrett constant and $\varepsilon_{B\infty}$ means the high-temperature permittivity. Parameters of the fits are summarized in Table 3. One can see that the extrapolated (hypothetical) critical temperature $T_{cr}$ increases and $T_1$ decreases with the Pb-concentration increase. However, the $T_1$ temperature, below which the quantum fluctuations start to play a role, is unusually high in our samples, much higher than in $SrTiO_3$, where $T_1$ is only ~60 K.[8]

Qualitatively different temperature dependence of IR reflectivity spectra and a higher number of polar phonons were observed in $Sr_{9-x}Pb_xCe_2Ti_{12}O_{36}$ samples for x≥3 in the FE phase. The lowest frequency phonon softens on cooling to $T_C$ and hardens on subsequent cooling. As an example, the IR reflectivity spectra of $Pb_9Ce_2Ti_{12}O_{36}$ at temperatures between 10 and 900 K are plotted in Fig. 12. One can see a dramatic change of the spectra shape with temperature due to the symmetry change. 3 broad reflection bands (fitted with 9 oscillators) are seen in the paraelectric phase at 900 K, while 15 oscillators were necessary for the fit at 10 K. The resulting complex permittivity obtained from the IR spectra fit is shown in Fig. 13 (the maxima in $\varepsilon''(\omega)$ spectra correspond roughly to phonon frequencies). The lowest-frequency phonon exhibits the most dramatic change with temperature (see Fig. 14). Its softening in the paraelectric phase fully explains the low frequency dielectric anomaly (Fig. 4). The phase transition is clearly of the displacive type. In the FE phase, the SM remarkably hardens from 13 (at 575 K) to 80 cm$^{-1}$ (at 10 K). Nevertheless, the low-frequency $\varepsilon'$ in the FE phase is larger than that given by the phonon contribution, obviously due to some contribution of the FE domain walls (see the discussion below).

**(c) Intrinsic and extrinsic MW dielectric loss**

More than twenty years ago, Wakino et al.[27,28] proposed the IR reflectivity spectroscopy as a tool for estimation of the intrinsic MW dielectric properties of MW materials. They used sum of damped harmonic oscillators[6] for modeling the dielectric dispersion in the IR frequency range:

$$\varepsilon^*(\omega) = \sum_{j=1}^{n} \frac{\Delta\varepsilon_j \omega_{TOj}^2}{\omega_{TOj}^2 - \omega^2 + i\omega\gamma_{TOj}} + \varepsilon_\infty \quad (6)$$

The parameters $\omega_{TOj}$ and $\gamma_{TOj}$ in Eq. (6) have the same meaning as in Eq. (2), $\Delta\varepsilon_j$ marks the contribution of the j-th phonon to the static permittivity. Actually, Eq. (6), which can be used in the case of a small splitting of longitudinal and transverse phonons, is a special case of the more general formula in Eq. (2).[6] Wakino et al.[27,28] suggested to use extrapolation of Eq.(6) from the IR down to the MW range, i.e. 2-3 orders of magnitude below phonon frequencies ($\omega \ll \omega_{TOj}$), which yields constant real part of the permittivity

$$\varepsilon' = \varepsilon_\infty + \sum_{j=1}^{n} \Delta\varepsilon_j , \qquad (7)$$

while the dielectric loss ε'' is proportional to the frequency ω

$$\varepsilon''(\omega) \propto \omega \sum_{j=1}^{n} \frac{\Delta\varepsilon \gamma_{TOj}}{\omega_{TOj}^2} . \qquad (8)$$

It implies that the dielectric loss could be linearly extrapolated from the THz range (0.1-3 THz) down to the MW region. Later on it was shown[29,30,31,4,5] that for an accurate determination of the complex permittivity in the THz range it is more suitable to combine the IR reflectivity with THz transmission spectroscopy, because the latter one is more sensitive to determine weak absorption processes below the strong one-phonon absorption peaks. Simultaneously, it was shown that the extrinsic absorption mechanisms only slightly contribute to the THz and IR absorption, therefore the linear extrapolation from the THz to the MW range allows us to estimate the intrinsic dielectric losses originating from the multiphonon absorption.[29,30,31,4]

The method described above was used for study of many MW ceramics[4,31] and it was shown that it gives a rather good estimate of the intrinsic MW losses, although Eq. (6) is valid only in the region near phonon frequencies and the use of this formula is not theoretically justified[1] for frequencies much lower than $\omega_{TOj}$, i.e. in the MW range. Gurevich and Tagantsev developed a comprehensive microscopic phonon transport theory[32] and have shown, that in the MW range the two-phonon difference decay processes dominate at room and medium-high temperatures and the theory predicts

$$\varepsilon''(\omega,T) \sim \omega T^2 . \qquad (9)$$

It means that both approaches, damped oscillator model as well as microscopic phonon transport theory yield the same linear frequency dependence of ε''(ω). Moreover, the microscopic theory predicts a quadratic temperature dependence of the dielectric loss.

Figs. 15 and 16 show the real and imaginary parts of ε* calculated from the IR and THz spectra fits extrapolated down to the MW region. Experimental MW and THz data are also plotted with the exception of x=6 sample, where high opacity prevented the THz measurements and only the result of IR spectra fit is shown. One can see that the MW values of ε' in $Sr_{9-x}Pb_xCe_2Ti_{12}O_{36}$ samples for x ≤ 4 are well described by phonon contributions, but in the samples with x ≥ 5 the extrapolated MW values of ε' are much lower than the experimental ones. This is apparently due to a weak MW relaxation arising probably from the FE domain walls contribution in the FE samples of x≥7 and from a possible Pb disorder in the paraelectric phase of the x = 5 sample (note that this sample is highly absorbing in THz range, therefore we obtained only rather inaccurate THz data).

Extrapolation of ε'' down to the MW region shows that the experimental MW values of ε'' for the 2 ≤ x ≤ 4 samples agree rather well with the extrapolated ones. This indicates that the samples are optimally processed and show almost no extrinsic dielectric loss. The loss spectra of x=0 and x=1 samples show a deviation of the theoretical ε''(ω) curves from the experimental points not only in the MW but also in the THz range. It indicates some extrinsic contribution to losses. In these cases the sample processing could be probably improved. The x ≥ 5 samples show much higher MW losses than those from the extrapolation, possibly due to the FE domain wall contribution which cannot be avoided in polydomain samples.

**4. Conclusions**

Novel $Sr_{9-x}Pb_xCe_2Ti_{12}O_{36}$ (x = 0-9) ceramic system was sintered and characterized in the broad frequency (100 Hz – 100 THz) and temperature (10 - 900 K) regions. The samples with low content of Pb (x ≤ 2) exhibit incipient FE behavior like the related $SrTiO_3$. Higher content of Pb induces a displacive FE phase transition with linearly increasing Curie-Weiss temperature $T_C$ with the Pb concentration. The refinement of the crystal structures by X-ray, SAED, IR and Raman data yield trigonal $R\bar{3}c$ space group for the paraelectric phase and monoclinic space group *Cc* for the FE phase. Optical soft phonon was observed in the IR and THz spectra of $Pb_9Ce_2Ti_{12}O_{36}$ so that the FE phase transition is of displacive type. Room-temperature MW dielectric properties of $Sr_{9-x}Pb_xCe_2Ti_{12}O_{36}$ (x = 0-4) ceramic are completely described by polar phonon contributions. The x ≥ 5 samples exhibit a MW relaxation due to the contribution of the FE domain walls (for x ≥ 7) and due to possible Pb anharmonic vibrations probably leading to crossover from displacive to order-disorder behavior (for the x=5 sample). The best MW properties are expected for $Sr_8PbCe_2Ti_{12}O_{36}$, although this ceramics was apparently not yet optimally processed and exhibited a large extrinsic dielectric loss.

**Acknowledgments**

The work was supported by the Czech Science Foundation (Project No. 202/06/0403) and by Czech Academy of Sciences (Project AVOZ10100520). G. Subodh is thankful to Council of Scientific and Industrial Research (CSIR), New Delhi, India, for the Senior Research Fellowship. We thank also to R. Ubic for a helpful discussion.

**Table 1.** The Rietveld refinement for ferroelectric Pb$_9$Ce$_2$Ti$_{12}$O$_{36}$ (space group *Cc*). The oxygen atoms were not refined, coordinates of Pb/Ce positions did not deviate significantly from their ideal positions. The isotropic displacement factors for oxygen atoms were constrained to be equal.

| Atom  | Site | x        | y        | z        | $B_{iso}$ |
|-------|------|----------|----------|----------|-----------|
| Pb,Ce | 4f   | 0.25     | 0.25     | 0.75     | 1.34(3)   |
| Ti    | 4f   | 0.520(1) | 0.243(2) | 0.515(2) | 1.2(1)    |
| O1    | 4f   | 0.006    | 0.993    | 0.763    | 1.6(1)    |
| O2    | 4f   | 0.75     | 0.76     | 0.25     | 1.6(1)    |
| O3    | 4f   | 0.99     | 0.49     | 0.736    | 1.6(1)    |

**Table 2.** Factor group analysis of optical lattice vibrations for the possible paraelectric and FE phases. IR and R means activity of the phonons in the IR and Raman spectra, respectively. (-) marks silent phonons inactive in any spectra. The 3 acoustic modes are subtracted.

In possible **paraelectric phases:**

$\Gamma_{Pm\bar{3}m} = 3F_{1u}(IR) + 1F_{2u}(-)$

$\Gamma_{I4/mcm} = 3A_{1u}(-) + 4A_{2u}(IR) + 8E_u(IR) + E_g(R) + B_{2g}(R) + B_{1u}(-)$

$\Gamma_{R\bar{3}c} = 3A_{2g}(-) + 3A_{2u}(IR) + 2A_{1u}(-) + A_{1g}(R) + 4E_g(R) + 5E_u(IR)$

In possible **FE phases:**

$\Gamma_{R3c} = 4A_1(IR,R) + 5A_2(-) + 9E(IR,R)$

$\Gamma_{Cc} = 15\ A'(IR,R) + 15A''(IR,R)$

**Table 3.** Parameters of the fit to the Barrett formula (Eq. 4) for the temperature dependence of low-frequency $\varepsilon'$ in $Sr_{9-x}Pb_xCe_2Ti_{12}O_{36}$ (x=0-2).

|  | x=0 | x=1 | x=2 |
| --- | --- | --- | --- |
| $T_1$ (K) | 185 | 121 | 139 |
| $T_{cr}$ (K) | -288 | -116 | -50 |
| $\varepsilon_{B\infty}$ | 28 | 20 | 18 |
| M (K) | 79800 | 90000 | 89500 |

**Figure captions**

**Figure 1.** X-ray diffraction patterns of the $Sr_{9-x}Pb_xCe_2Ti_{12}O_{36}$ ceramics.

**Figure 2.** Selected-area electron diffraction patterns of $Sr_9Ce_2Ti_{12}O_{36}$ corresponding to (a)[-111] direction (note the satellites), and $Pb_9Ce_2Ti_{12}O_{36}$ corresponding (b) [011], (c) [010] and (d) [103] directions showing the pseudotetragonal relationship of the $Pb_9Ce_2Ti_{12}O_{36}$ phase.

**Figure 3**. Rietveld refinement of $Pb_9Ce_2Ti_{12}O_{36}$ ceramics in the space group *Cc*. The inset shows unsuccessful refinement in the space group *R3c*. Observed data are shown by circles, calculated and differences profiles are represented by solid lines. The upper reflections bars correspond to $Pb_9Ce_2Ti_{12}O_{36}$, the middle and lower bars to 3 and 2 mass percent of $CeO_2$ and $TiO_2$ impurity, respectively.

**Figure 4.** Temperature dependences of the real and imaginary parts of complex permittivity for the $Sr_{9-x}Pb_xCe_2Ti_{12}O_{36}$ (x = 0-9) ceramics. Data were taken at 100 kHz (x = 0-3, 9) and at 100 MHz (x = 4-8).

**Figure 5.** Phase diagram for the $Sr_{9-x}Pb_xCe_2Ti_{12}O_{36}$ (x = 0-9) ceramics. The inset shows the FE hysteresis loop taken at 10 Hz and 300 K from the $Pb_9Ce_2Ti_{12}O_{36}$ ceramics.

**Figure 6.** Concentration dependence of the MW permittivity and Qxf parameters for the $Sr_{9-x}Pb_xCe_2Ti_{12}O_{36}$ (x = 0-9) ceramics. Data were collected at room temperature between 0.9 and 3.2 GHz. In some cases several samples of the same chemical composition were measured differing in size (see several data for the same x).

**Figure 7.** Room-temperature IR reflectivity of selected $Sr_{9-x}Pb_xCe_2Ti_{12}O_{36}$ ceramics.

**Figure 8.** Room-temperature micro-Raman spectra of $Sr_{9-x}Pb_xCe_2Ti_{12}O_{36}$ ceramics (x=0-9). The peak at 465 cm$^{-1}$ is attributed to the $CeO_2$ second phase.

**Figure 9.** Complex dielectric spectra of $Sr_{9-x}Pb_xCe_2Ti_{12}O_{36}$ ceramics (x=0-9) obtained from the fit of room-temperature IR reflectivity shown in Fig. 6. The full symbols are THz data. The lowest frequency peak in the $\varepsilon''$ spectra corresponds to the FE soft mode.

**Figure 10.** Room-temperature transverse-phonon frequencies vs Pb concentration in $Sr_{9-x}Pb_xCe_2Ti_{12}O_{36}$ ceramics. The inset shows the dependence of the FE soft mode frequency. Note the appearance of new modes in x ≥ 7 compounds in the FE phase.

**Figure 11.** Temperature dependence of the IR reflectivity spectra for the $Sr_9Ce_2Ti_{12}O_{36}$ ceramics. The spectra are plotted at each 50 K below 300 K. Inset shows the temperature dependence of permittivity at 1 MHz (solid line), dots mark static permittivity obtained from IR spectra fit. Note the saturation of ε'(T) at low temperatures due to quantum fluctuations.

**Figure** 12. Temperature dependence of the IR reflectivity for the $Pb_9Ce_2Ti_{12}O_{36}$ ceramics.

**Figure 13.** Temperature dependence of the complex dielectric response in $Pb_9Ce_2Ti_{12}O_{36}$ ceramics obtained from the IR reflectivity and THz data (full symbols) fits. The sample was almost opaque near $T_C$=550 K, therefore the THz data at 575 K are less accurate than the rest of the spectra. THz spectra were not measured at 10 K. The deviations of the experimental ε'' THz spectra from the fitted ones at 300 and 500 K are due to extrinsic contributions.

**Figure 14.** Temperature dependence of the two lowest-frequency polar modes in $Pb_9Ce_2Ti_{12}O_{36}$ ceramics. FE soft mode shows a clear anomaly near $T_C$ = 550 K.

**Figure 15.** Real part of the complex permittivity calculated from the room-temperature IR spectra fits and extrapolated down to the MW range. Black solid squares mark experimental THz data, red solid symbols below 0.1 cm$^{-1}$ (i.e. 3 GHz) mark experimental MW ε' data.

**Figure 16.** Imaginary part of the complex permittivity of $Sr_{9-x}Pb_xCe_2Ti_{12}O_{36}$ (x=0-9) ceramics calculated from the room-temperature IR spectra fits and extrapolated down to the MW range. Black solid squares mark the THz data, red solid symbols below 0.1 cm$^{-1}$ (i.e. 3 GHz) mark the MW ε'' data.

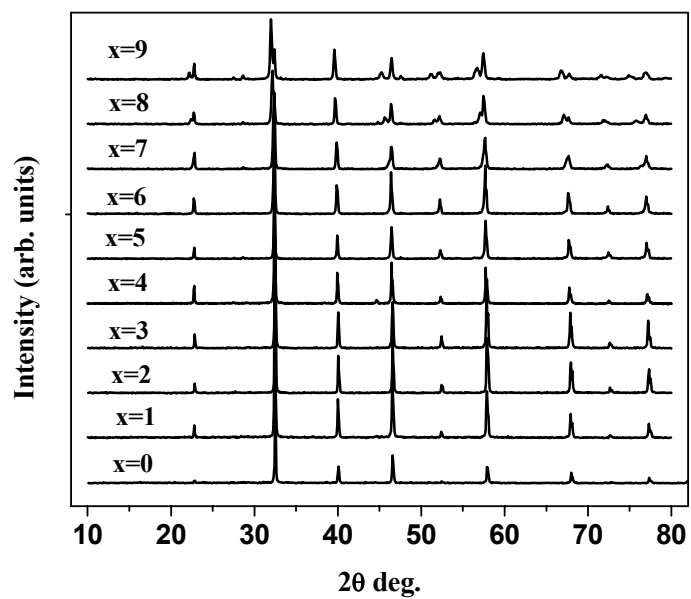

**Figure 1.** X-ray diffraction patterns of the $Sr_{9-x}Pb_xCe_2Ti_{12}O_{36}$ ceramics.

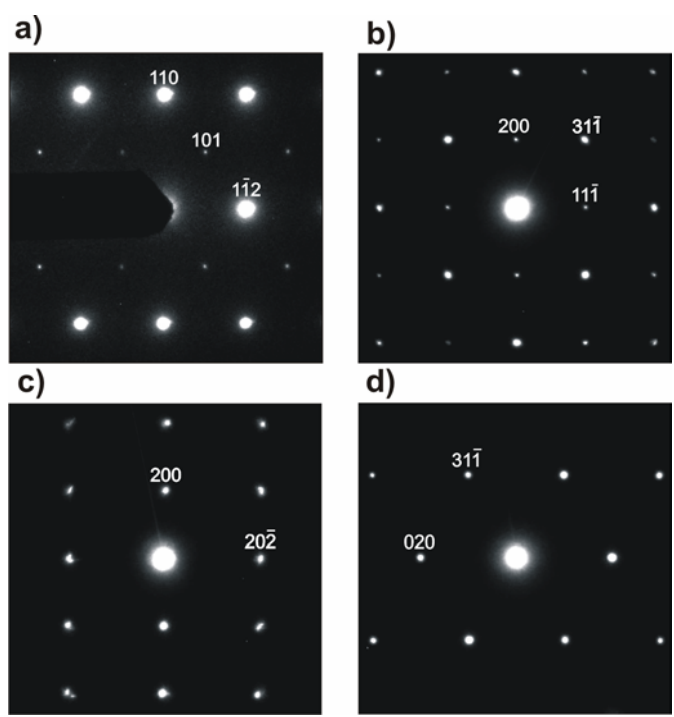

**Figure 2.** Selected-area electron diffraction patterns of $Sr_9Ce_2Ti_{12}O_{36}$ corresponding to (a)[-111] direction (note the satellites), and $Pb_9Ce_2Ti_{12}O_{36}$ corresponding (b) [011], (c) [010] and (d) [103] directions showing the pseudotetragonal relationship of the $Pb_9Ce_2Ti_{12}O_{36}$ phase.

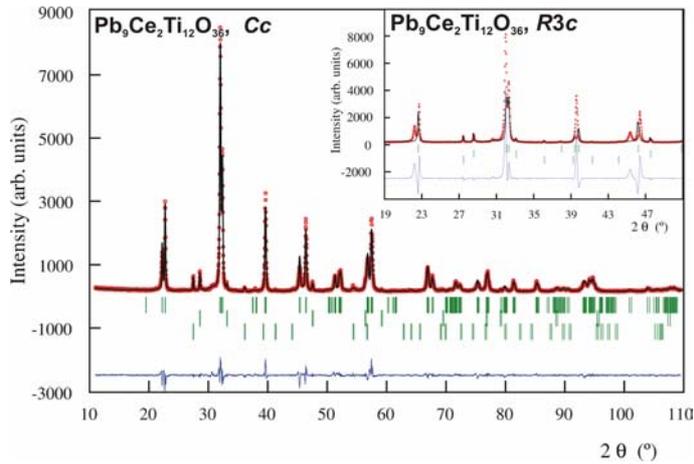

**Figure 3.** Rietveld refinement of $Pb_9Ce_2Ti_{12}O_{36}$ ceramics in the space group *Cc*. The inset shows unsuccessful refinement in the space group *R3c*. Observed data are shown by circles, calculated and differences profiles are represented by solid lines. The upper reflections bars correspond to $Pb_9Ce_2Ti_{12}O_{36}$, the middle and lower bars to 3 and 2 mass percent of $CeO_2$ and $TiO_2$ impurity, respectively.

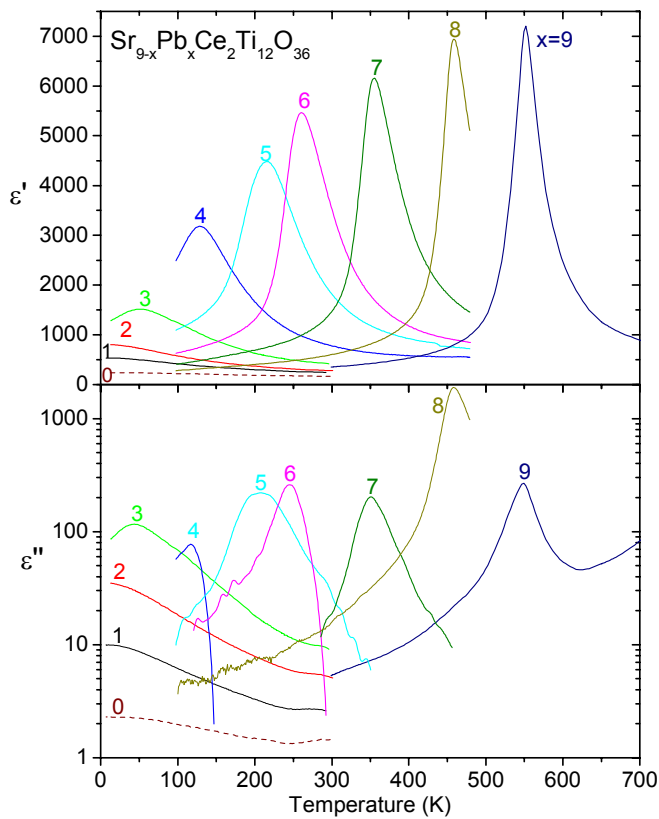

**Figure 4.** Temperature dependences of the real and imaginary parts of complex permittivity for the $Sr_{9-x}Pb_xCe_2Ti_{12}O_{36}$ (x = 0-9) ceramics. Data were taken at 100 kHz (x = 0-3, 9) and at 100 MHz (x = 4-8).

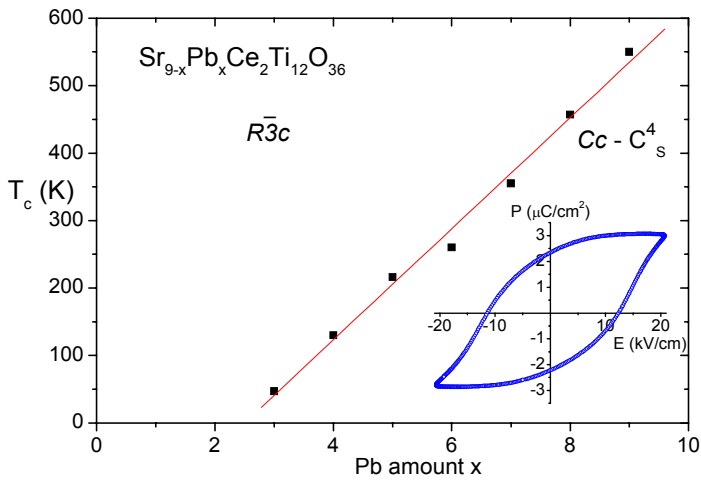

**Figure 5.** Phase diagram for the $Sr_{9-x}Pb_xCe_2Ti_{12}O_{36}$ (x = 0-9) ceramics. The inset shows the FE hysteresis loop taken at 10 Hz and 300 K from the $Pb_9Ce_2Ti_{12}O_{36}$ ceramics.

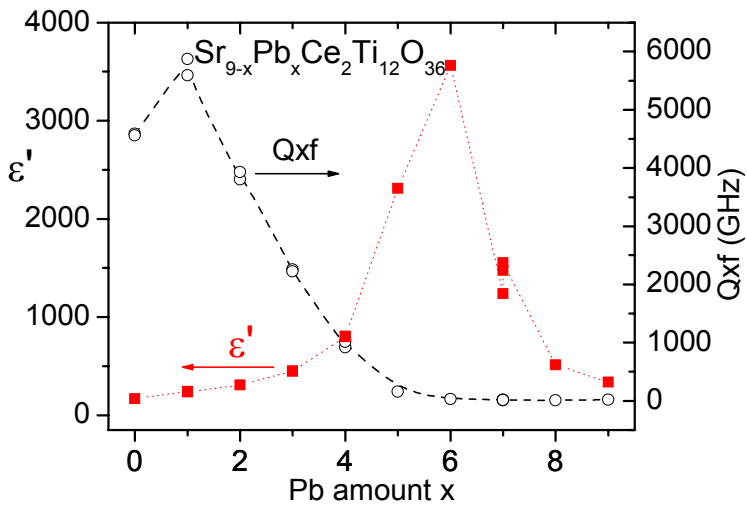

**Figure 6.** Concentration dependence of the MW permittivity and Qxf parameters for the $Sr_{9-x}Pb_xCe_2Ti_{12}O_{36}$ (x = 0-9) ceramics. Data were collected at room temperature between 0.9 and 3.2 GHz. In some cases several samples of the same chemical composition were measured differing in size (see several data for the same x).

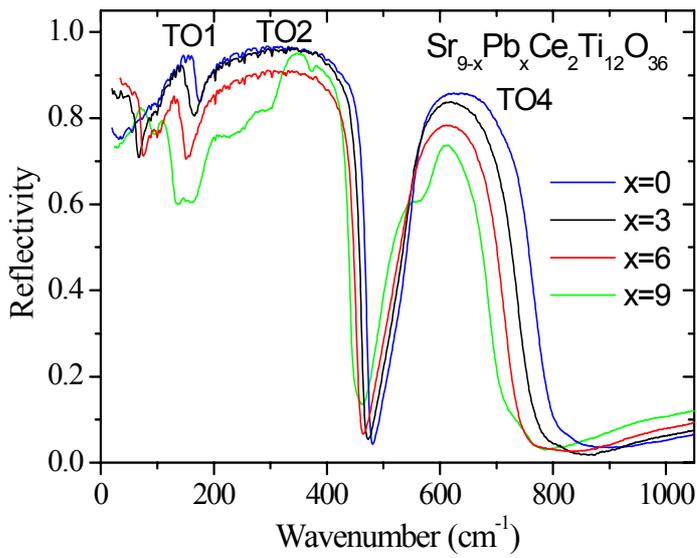

**Figure 7.** Room-temperature IR reflectivity of selected $Sr_{9-x}Pb_xCe_2Ti_{12}O_{36}$ ceramics.

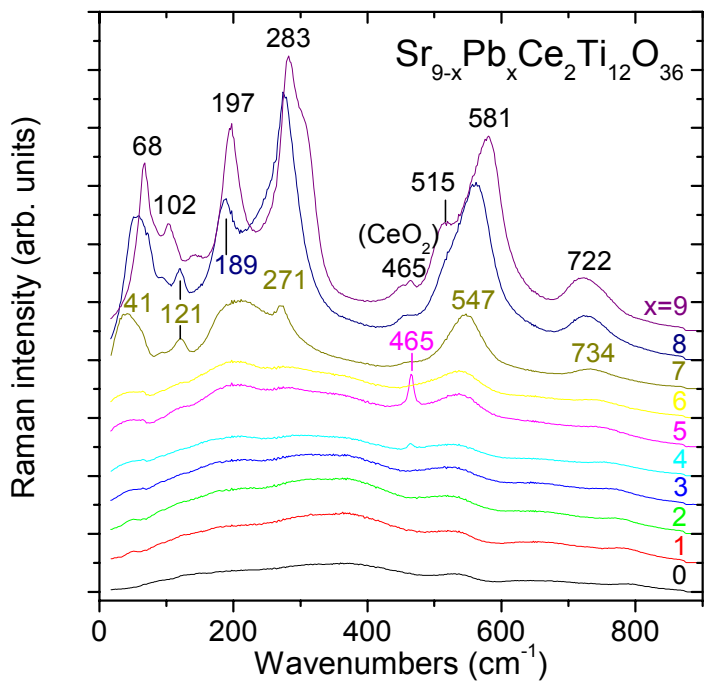

**Figure 8.** Room-temperature micro-Raman spectra of $Sr_{9-x}Pb_xCe_2Ti_{12}O_{36}$ ceramics (x=0-9). The peak at 465 cm$^{-1}$ is attributed to the $CeO_2$ second phase.

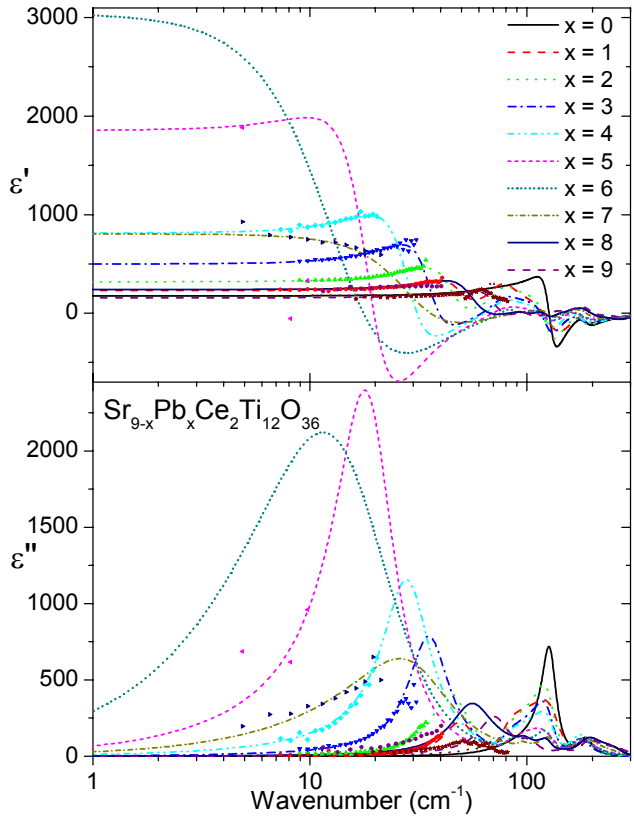

**Figure 9.** Complex dielectric spectra of $Sr_{9-x}Pb_xCe_2Ti_{12}O_{36}$ ceramics (x=0-9) obtained from the fit of room-temperature IR reflectivity shown in Fig. 6. The full symbols are THz data. The lowest frequency peak in the ε'' spectra corresponds to the FE soft mode.

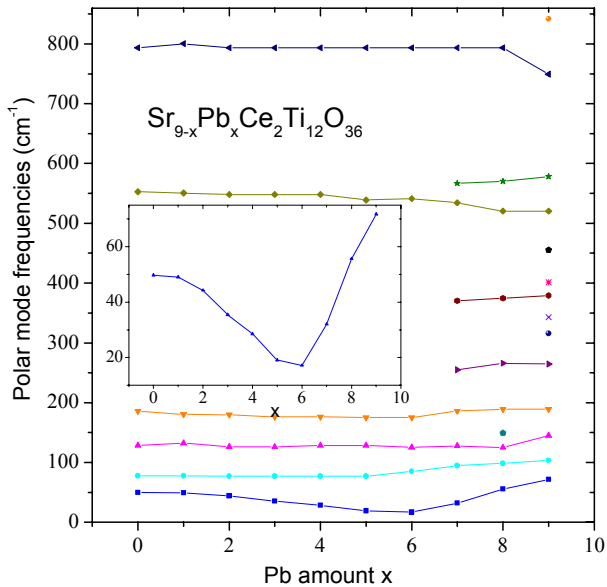

**Figure 10.** Room-temperature transverse-phonon frequencies vs Pb concentration in $Sr_{9-x}Pb_xCe_2Ti_{12}O_{36}$ ceramics. The inset shows the dependence of the FE soft mode frequency. Note the appearance of new modes in x ≥ 7 compounds in the FE phase.

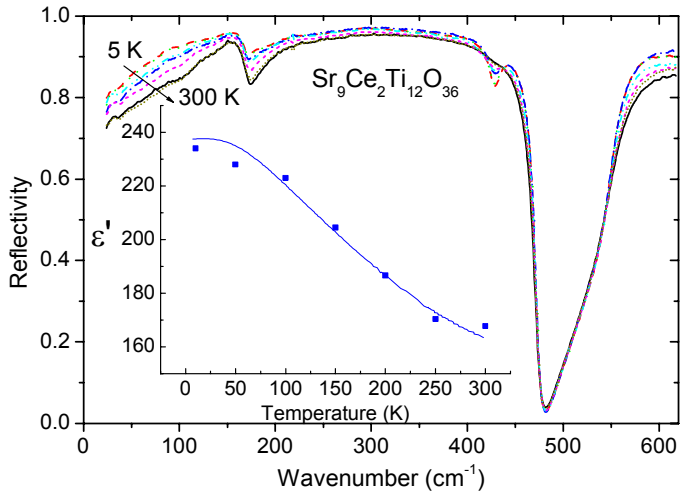

**Figure 11.** Temperature dependence of the IR reflectivity spectra for the $Sr_9Ce_2Ti_{12}O_{36}$ ceramics. The spectra are plotted at each 50 K below 300 K. Inset shows the temperature dependence of permittivity at 1 MHz (solid line), dots mark static permittivity obtained from IR spectra fit. Note the saturation of $\varepsilon'(T)$ at low temperatures due to quantum fluctuations.

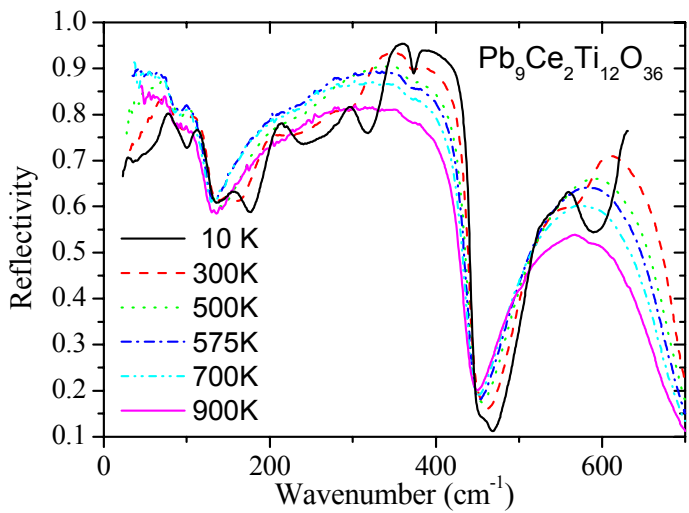

**Figure 12.** Temperature dependence of the IR reflectivity for the $Pb_9Ce_2Ti_{12}O_{36}$ ceramics.

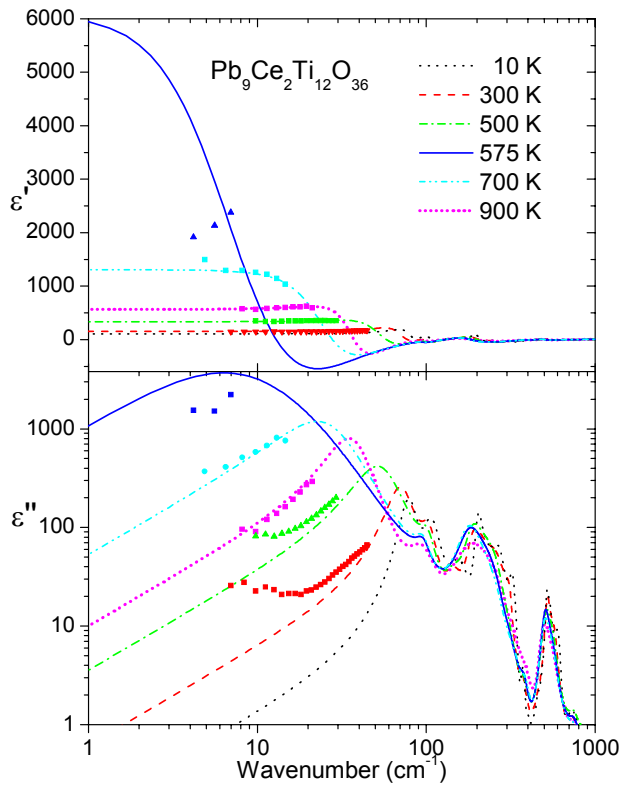

**Figure 13.** Temperature dependence of the complex dielectric response in $Pb_9Ce_2Ti_{12}O_{36}$ ceramics obtained from the IR reflectivity and THz data (full symbols) fits. The sample was almost opaque near $T_C$=550 K, therefore the THz data at 575 K are less accurate than the rest of the spectra. THz spectra were not measured at 10 K. The deviations of the experimental ε'' THz spectra from the fitted ones at 300 and 500 K are due to extrinsic contributions.

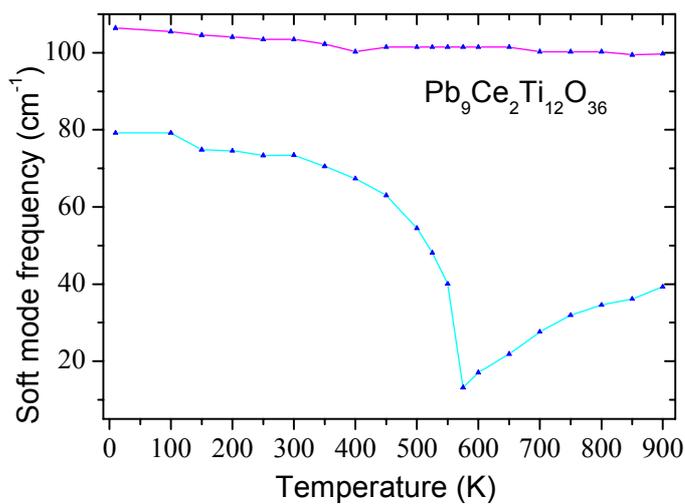

**Figure 14.** Temperature dependence of the two lowest-frequency polar modes in $Pb_9Ce_2Ti_{12}O_{36}$ ceramics. FE soft mode shows a clear anomaly near $T_C$ = 550 K.

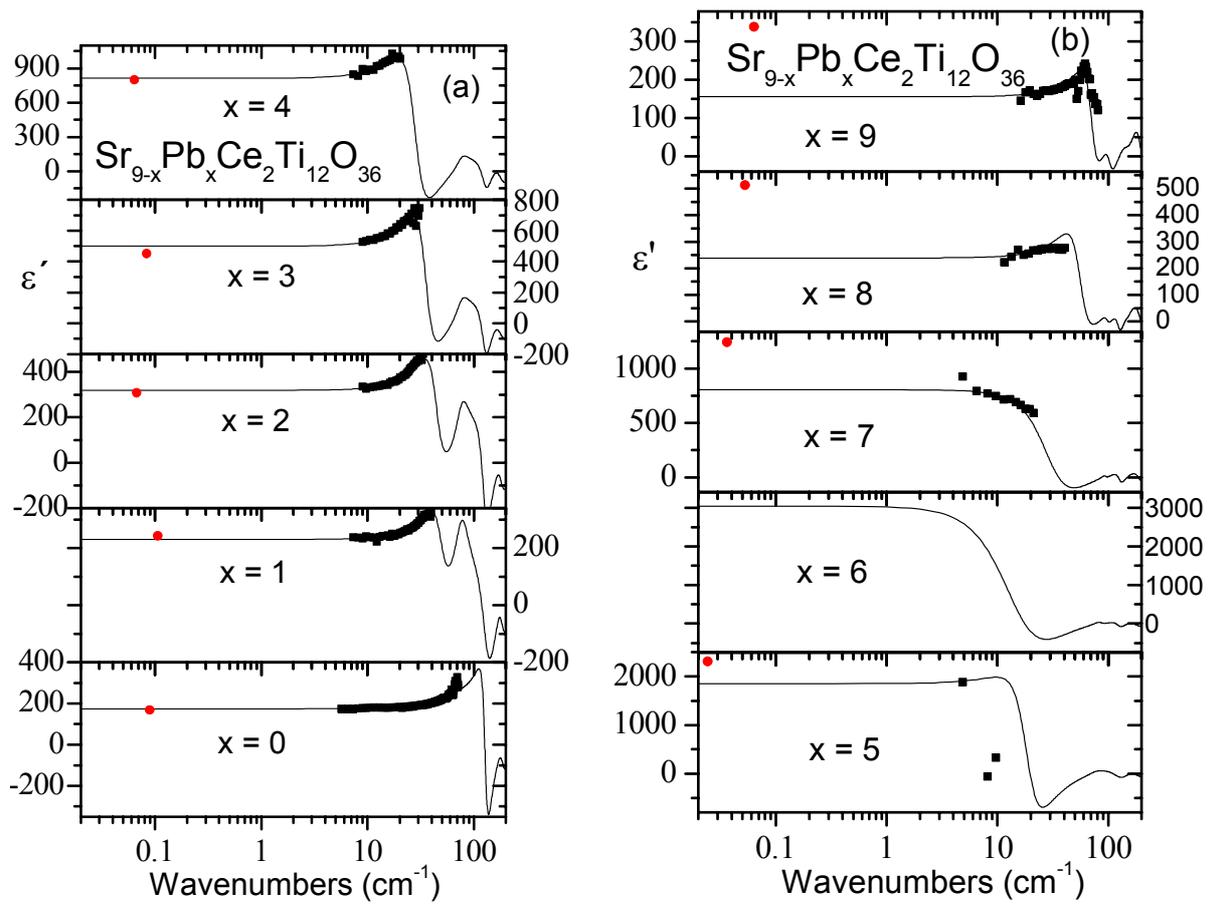

**Figure 15.** Real part of the complex permittivity calculated from the room-temperature IR spectra fits and extrapolated down to the MW range. Black solid squares mark experimental THz data, red solid symbols below 0.1 cm$^{-1}$ (i.e. 3 GHz) mark experimental MW $\varepsilon'$ data.

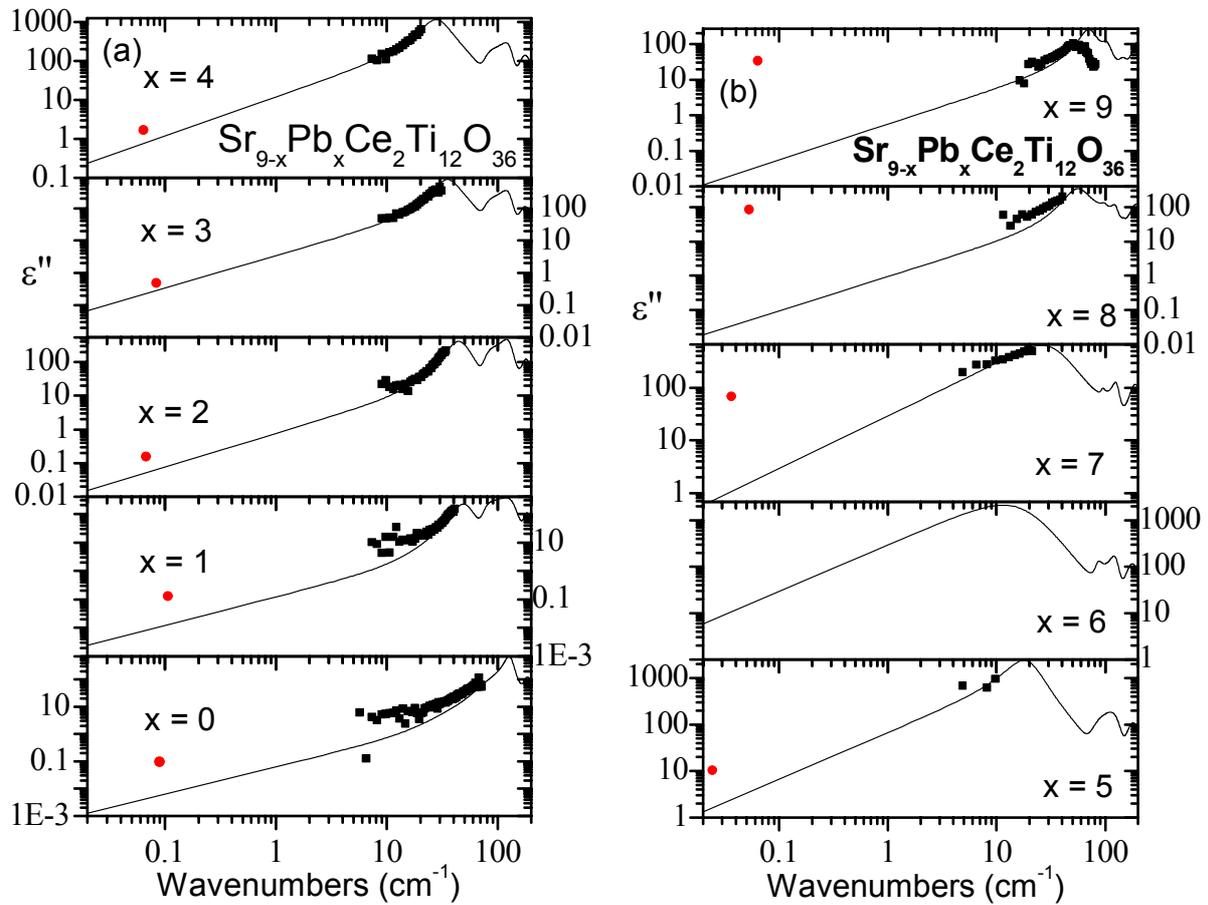

**Figure 16.** Imaginary part of the complex permittivity of $Sr_{9-x}Pb_xCe_2Ti_{12}O_{36}$ (x=0-9) ceramics calculated from the room-temperature IR spectra fits and extrapolated down to the MW range. Black solid squares mark the THz data, red solid symbols below 0.1 cm$^{-1}$ (i.e. 3 GHz) mark the MW $\varepsilon''$ data.